\documentclass[pdflatex,twocolumn,epjc3]{svjour3}
\RequirePackage[T1]{fontenc}

\smartqed  

\RequirePackage{graphicx}
\RequirePackage{mathptmx}      
\RequirePackage{flushend}
\RequirePackage[numbers,sort&compress]{natbib}
\RequirePackage[colorlinks,citecolor=blue,urlcolor=blue,linkcolor=blue]{hyperref}

\journalname{Eur. Phys. J. C}

\begin{document}
\title{Search for axioelectric effect of 5.5 MeV solar axions using BGO detectors}

\author{A.V.~Derbin, S.V.~Bakhlanov, I.S.~Dratchnev, A.S.~Kayunov, V.N.~Muratova}

\institute{St.Petersburg Nuclear Physics Institute, Gatchina, Russia 188300}
\date{Received: date / Accepted: date}
\maketitle
\begin{abstract}
A search for axioelectric absorption of solar axions produced in the $ p + d \rightarrow {^3\rm{He}}+\gamma~(5.5~
\rm{MeV})$  reactions has been performed with a BGO detector placed in a low-background setup. A model-independent
limit on an axion-nucleon and axion-electron coupling constant has been obtained: $| g_{Ae}\times g_{AN}^3|< 2.9\times
10^{-9}$ for 90\% confidence level. The constrains of the axion-electron coupling have been obtained for hadronic axion
with masses in (0.1 - 1) MeV range: $|g_{Ae}| \leq (1.4 - 9.7)\times 10^{-7}$.
\end{abstract}

\section{INTRODUCTION}

The axion is a pseudo-Goldstone boson arising from breaking the global Peccei-Quinn (PQ) symmetry
\cite{Pec77,Wei78,Wil78}.  The original so-called Weinberg-Wilczek-Peccei-Quinn axion model contained some certain
direct predictions for the coupling constants between axions and photons ($ g_{A\gamma}$), electrons ($ g_{Ae}$) and
nucleons ($ g_{AN}$) as well as for the axion mass ($m_A$). This model was quickly disproved by the reactor and
accelerator experiments \cite{PDG12}.

The particle has been retained by two new theoretical models of the "invisible" axion in the form required  to solve
the CP problem of strong interactions; at the same time its interaction with matter is suppressed. These are models of
the "hadronic"(or Kim-Shifman-Vainstein-Zakharov (KSVZ)) \cite{Kim79,Shi80} and the "GUT" (or Dine-Fischler-Srednicki-
Zhitnitskii (DFSZ)) \cite{Zhi80,Din81} axion. The scale of Peccei-Quinn symmetry violation is arbitrary in both models
and can be extended up to the Planck mass $\approx 10^{19}~ \rm{GeV}$. The axion mass in these models is determined by
the axion decay constant $f_A$:
\begin{equation}
  m_A\approx \frac{f_\pi m_\pi\sqrt{z}}{f_A(1+z)};~~~m_A {\rm{(eV)}} \approx \frac{6.0\times10^6}{f_A\rm{(GeV)}},
\end{equation}
where $m_\pi$ and $f_\pi$ are, respectively, the mass and decay constant of neutral $\pi$-meson and   $z = m_u/m_d$ is
quark-mass ratio. Since the axion coupling constants $g_{A\gamma}$, $g_{Ae}$ and $g_{AN}$ are proportional to the axion
mass, the interaction of an axion with matter is suppressed.

The main experimental efforts are focused on searches for an axion with a mass in the range of  $10^{-6}$ to $10^{-2}$
eV. As this range is free of astrophysical and cosmological constraints, relic axions with such a mass could be
considered to be the most likely candidates for dark matter particles. Experimental bounds on the mass of the axion
follow from the constraints on the $g_{A\gamma}$, $g_{Ae}$, and $g_{AN}$ coupling constants, which significantly depend
on the theoretical model used.

New possibilities for solving a strong CP problem are based on the concept of the existence of the  mirror particles
\cite{Ber01} and supersymmetry \cite{Hal04}. These models suppose the existence of axions with the mass of about 1 MeV,
and this existence is forbidden by neither laboratory experiments nor astrophysical data.

This work is devoted to the search for solar axions with an energy of 5.5 MeV,  produced in the  $p + d
\rightarrow\rm{^3He}+ A$  reaction. The range of axion masses under study is expanded up to 5 MeV. The axion flux is
proportional to the $pp$-neutrino flux, which is estimated with a high degree of accuracy \cite{Ser09}. The axion
interaction exploited in this study is the axioelectric effect ${\rm A}+e+Z\rightarrow e+Z$. As the cross section of
the reaction depends on the charge of a nucleus as $Z^5$, the BGO detector containing bismuth (stable nucleus with the
largest charge $Z=83$) is the most suitable from this point of view.

Recently, the high energy solar axions and axions from a nuclear reactor have been sought by the Borexino
\cite{Bel08,Bel12}, the CAST \cite{And10} and the Texono \cite{Cha07} collaborations. We have previously published a
search for 5.5 MeV axions with two small BGO detectors \cite{Der10}.

\section{ AXION PRODUCTION IN NUCLEAR MAGNETIC TRANSITIONS AND THE AXIOELECTRIC EFFECT}
Since the temperature in the center of the Sun is about 1.3 keV, the Sun should be an intense source  of low energy
$\sim$(1--10) keV axions.
The reactions of the main solar chain and $CNO$ cycle can produce axions with higher energies. The most intensive flux
is expected as a result of the reaction: $p + d \rightarrow {^3\rm{He}} + \gamma$ when 5.5 MeV axion is emitted instead
$\gamma-$quantum.

According to the standard solar model (SSM), 99.7\% of all deuterium is produced as a result of the two protons
fusion, $p + p \rightarrow d + e^+ + \nu_e$, while the remaining 0.3\% is due to the  $p+ p + e^- \rightarrow  d +
\nu_e$ reaction. The expected solar axion flux can thus be expressed in terms of the $pp$-neutrino flux, which is
$6.0\times 10^{10} {\rm{cm}}^{-2} {\rm{s}}^{-1}$ \cite{Ser09}.  The proportionality factor between the axion and
neutrino fluxes is determined by the axion-nucleon coupling constant $g_{AN}$, which consists of isoscalar $g^0_{AN}$
and  isovector $g^3_{AN}$ components.

In the $p(d,{^3\rm{He}})\gamma$ reaction, the M1-type transition corresponds to the capture of a proton  with a zero
orbital momentum. The probability of proton capture from the $S$ state at proton energies below 80 keV has been
measured in \cite{Sch97}; at the proton energy of $\sim 1$ keV, M1 fraction of the total $p(d,{^3\rm{He}})\gamma$ cross
section is $\chi$ = 0.55. The proton capture from the $S$ state corresponds to the isovector transition, and the  ratio
of the probability of a nuclear transition with the axion production $(\omega_A)$ to the probability of a magnetic
transition $(\omega_\gamma)$ depends only on $g^3_{AN}$ \cite{Don78,Raf82,Avi88,Hax91,Der97}:
\begin{equation}\label{ratio}
\frac{\omega_{A}}{\omega_{\gamma}} =
 \frac{\chi}{2\pi\alpha}\left[\frac{g_{AN}^{3}}{\mu_3}\right]^2\left(\frac{p_A}{p_\gamma}\right)^3 = 0.54(g_{AN}^{3})^2
 \left(\frac{p_A}{p_\gamma}\right)^3.
\end{equation}
where $p_{\gamma}$ and $p_{A}$ are, respectively, the photon and axion momenta; $\alpha\approx 1/137$ is  the
fine-structure constant;  and $\mu_3 = \mu_p - \mu_n \approx 4.71$ is isovector nuclear magnetic momenta.

Within the hadronic axion model, the constant $g^3_{AN}$ can be written in terms of the axion mass \cite{Sre85, Kap85}:
\begin{equation}\label{gan3}
g_{AN}^{3}=-2.75 \times 10^{-8}(m_A/1 {\rm{eV}}).
\end{equation}

The value of $g^3_{AN}$ in the DFSZ model depends on an additional unknown parameter $cos^2\beta$, but  have the same
order of magnitude. The numerical value lies in the range of ($0.3 - 1.5$) from the the value of $g^3_{AN}$ for the
hadronic axion \cite{Sre85}.

\begin{figure}
\includegraphics[width=9cm,height=10.5cm]{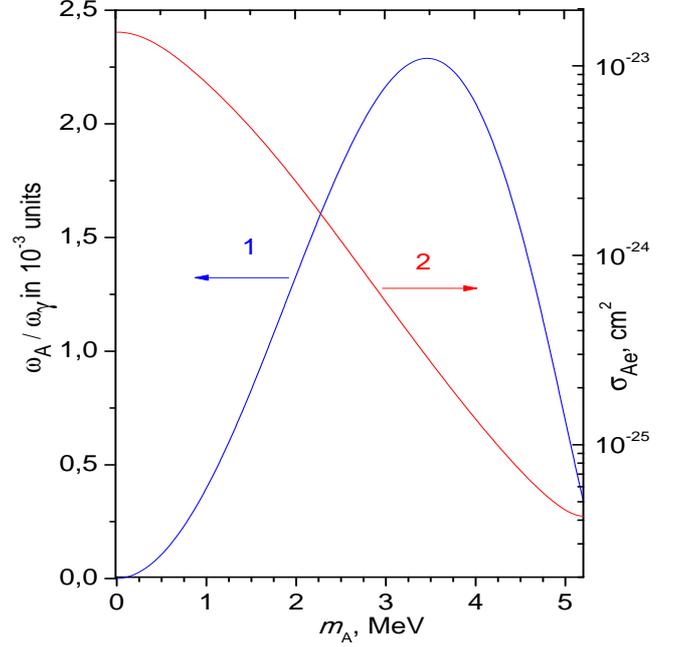}
\caption {The ratio of the emission probabilities for axions and $\gamma$ quanta ($\omega_A/\omega_\gamma$)  in the $p
+ d \rightarrow {^3{\rm{He}}} + \gamma$ reaction (curve 1, left-hand scale); the cross section of the axioelectric
effect for 5.5-MeV axions on bismuth atoms for $g_{Ae} = 1$ (curve 2, right-hand scale). } \label{fig1}
\end{figure}

The calculated values of the $\omega_A/\omega_\gamma$ ratio as a function of the axion mass are shown in
Fig.\ref{fig1}.  The axion flux on the Earth's surface is
\begin{eqnarray}\label{FluxA}
\Phi_A = \Phi_{\nu p p}(\omega_A/\omega_\gamma)
\end{eqnarray}
where $\Phi_{\nu p p} = 6.0 \times 10^{10} {\rm{cm}}^{-2} {\rm{s}}^{-1}$ is the $pp$-neutrino flux.

To detect 5.5 MeV axions, we chose the reaction of axioelectric effect $A + Z + e \rightarrow Z + e$   which is caused
by the axion-electron interaction.  The cross section of the axioelectric effect depends on the nuclear charge
according to the $Z^5$ law, and therefore, it is  reasonable to search for this process using detectors with a large
$Z$. The cross section for bismuth atoms exceeds the cross section of the Compton conversion of an axion ($A +e^-
\rightarrow\gamma + e^-$) by almost two orders of magnitude. The detection efficiency for the produced electron is
almost 100\% and the background level at 5.5 MeV is much lower than in the range of natural radioactivity. As a result,
the sensitivity to constants $g_{Ae}$ and $g_{AN}$ can be high even in an experiment using a relatively small target
mass.

In the axioelectric effect (an analog of the photoelectric effect), an axion disappears and an electron  with the
energy of $E_e=E_A - E_b$, where $E_b$ is the electron binding energy, is emitted from the atom. The axioelectric
effect cross section for K-shell electrons has been calculated (on the assumption that $E_A\gg E_b$ and $Z\ll 137$) in
\cite{Zhi79}:
\begin{center}
\begin{eqnarray}
\nonumber \sigma_{Ae} = 2(Z\alpha m_e)^5\frac{g^2_{Ae}}{m_e^2}\frac{p_e} {p_A}\ [\frac{4E_A(E^2_A+m^2_A)}{(p^2_A- p^2_e)^4}-\frac{2E_A}{(p^2_A -p^2_e)^3} \\
\nonumber-\frac{64}{3}p^2_ep^2_Am_e\frac{m^2_A}{(p^2_A -p^2_e)^6}-\frac{16m^2_Ap^2_AE_e}{(p^2_A-p^2_e)^5}-~~\\
-\frac{E_A}{p_ep_A}\frac{1}{(p^2_A-p^2_e)^2}\ln\frac{p_e+p_A}{p_e-p_A}].~~~\label{sigmaAE}
\end{eqnarray}
\end{center}

The dependence of the cross section on the axion mass for the coupling constant $g_{Ae} = 1$ is shown in  Fig. 1. The
K-shell electrons make the main contribution to the cross section. The contribution from the other electrons was
incorporated by introducing a factor of $5/4$, by analogy with the photoelectric effect.

\section{ INTERACTION OF AXIONS WITH SOLAR MATTER AND AXION DECAYS}

The flux of 5.5 MeV axions on the Earth's surface is proportional to the $pp$-neutrino flux only when the   axion
lifetime exceeds the time of a flight from the Sun and when the flux is not reduced as a result of the axion absorption
by matter inside the Sun. The requirement that most axions escape the Sun and reach the Earth thus limits the axion
coupling strengths accessible to terrestrial experiments \cite{Raf82,Bel08,Der10,Bel12}.

Axions leaving the center of the Sun pass through the matter layer of $\approx 6.8\times 10^{35}$ electrons and
$\approx 5\times 10^{35}$ protons per ${\rm{cm}}^{-2}$ in order to reach the Sun's surface. The maximum cross section
of the axio-electric effect on atoms for 5.5 MeV axions is $\sigma_{Ae}\approx g^2_{Ae} Z^5 3.8\times
10^{-33}{\rm{cm}}^2$ (see Fig.\ref{fig1} for Bi).  Axion loss due to axioelectric absorption by Fe atoms (Z=26,
$N_{\rm{Fe}}/N_{\rm{H}} = 2.8\times 10^{-5}$) imposes an upper limit on $|g_{Ae}| < 10^{-4}$ above which the
sensitivity of terrestrial experiments using solar axions is reduced. The abundance of heavy ($Z > 50$) elements in the
Sun is $N_{\rm{Z}}/N_{\rm{H}}\sim10^{-9}$ in relation to hydrogen \cite{Asp06}. If $|g_{Ae}|< 10^{-3}$, the change in
the axion flux does not exceed 10\%.

The other process - Compton conversion of an axion into a photon - imposes stronger limit on the sensitivity of
Earth-bound  experiments to the constant $g_{Ae}$. The cross section of this reaction for 5.5-MeV axions depends weakly
on the axion mass and can be written as $\sigma_{cc}\approx g^2_{Ae}4\times 10^{-25} {\rm{cm}}^2$ \cite{Don78,Bel12}.
For $|g_{Ae}|$ values below $10^{-6}$, the axion flux is not substantially suppressed.

The axion-photon interaction, as determined by the constant $g_{A\gamma}$, leads to the conversion of an axion  into a
photon in a field of nucleus (Primakoff conversion). The cross section of the reaction is
$\sigma_{pc}(5.5\rm{MeV})\approx g^2_{A\gamma}Z^2 2\times 10^{-29}{\rm{cm}}^2$. Taking into account the density of
protons and $^4\rm{He}$ nuclei, the condition that axions efficiently escape the Sun imposes the constraint
$|g_{A\gamma}| < 10^{-4}~{\rm{GeV}}^{-1}$. Constraint for the other elements are negligible due to their low
concentration in the Sun.

The axion-nucleon interaction leads to axion absorption in a threshold reaction similar to photo--dissociation:  $A+ Z
\rightarrow Z_1 + Z_2$.  For axions with energy 5.5-MeV this can occur for only a few nuclei: $^{17}{\rm{O}}$,$
^{13}{\rm{C}}$, $^3\rm{He}$ and $^2{\rm{H}}$ \cite{Raf82}. It was shown in \cite{Raf82} that the absorption of axions
can be mainly due to the process $A + ^{17}\rm{O}\rightarrow ^{16}\rm{O} + n$. The cross section of the reaction
depends on the combination of iso-scalar and iso-vector coupling constants. If $|-g_{AN}^0+g_{AN}^3|< 10^{-2}$, the
change in the axion flux does not exceed 10\%. Because the absorption of axions due to isovector-transition
$A(^3\rm{He},d)p$ is insignificant, the range of values $|g_{AN}^3|$ available for the study is practically unlimited.

For the axions with a mass above $2m_e$, the main decay mode is the decay into an electron-positron pair.  The
condition that $90\%$ of all axions reach the Earth limits the sensitivity of the solar axion experiments to $|g_{Ae}|
< (10^{-12}-10^{-11})$ \cite{Der10}. On the other hand the measured value of the interplanetary positron flux allowed
us to establish a new constraint on the axion-electron coupling constant: $g_{Ae}\leq (1-5) \times 10^{-17}$ for axions
with masses in the range of ($1.2-5.4$) MeV \cite{Der10}.

If the axion mass is less than $2m_e$, $A\rightarrow e^+ + e^-$ decay is forbidden, but the axion can decay into two
$\gamma$ quanta. The probability of the decay depends on the axion-photon coupling constant and the axion mass -
$\tau_{A\rightarrow\gamma\gamma}=64\pi/g^2_{A\gamma}m^3_A$. The present-day experimental constraint on $g_{A\gamma}$ is
$10^{-9}$ ${\rm{GeV}}^{-1}$, which corresponds to $\tau_{cm} = 10^5$ s for 1-MeV axions. This means that the axion flux
is not practically reduced due to the $A\rightarrow 2\gamma$ decay even for axions with 5 MeV mass.

\section{EXPERIMENTAL SETUP}
We used a 2.46 kg BGO crystal, manufactured from bismuth orthogermanate ${\rm{Bi}}_4{\rm{Ge}}_3{\rm{O}}_{12}$ (1.65 kg
of Bi),  to search for the 5.5 MeV axions. The BGO crystal was grown at the Nikolaev Institute of Inorganic Chemistry
and it was shaped as a cylinder, 76 mm in diameter and 76 mm in height. The detector signal was measured by an R2887
photoelectron multiplier, which had an optical contact with a crystal end surface.

The spectrometric channel of the BGO scintillation detector included an amplifier with the shaping time of  1 $\mu$s
and a 12-digit ADC. The amplification was selected so that the ADC channel scale was 2.7 keV. The standard calibration
sources ($^{60}{\rm{Co}}$ and $^{207}{\rm{Bi}}$), in combination with the natural radioactivity lines of
$^{40}{\rm{K}}$ and the uranium and thorium families, were used for the energy calibration  of the detector. The energy
dependence of the detector resolution $\sigma$ can be presented as $\sigma/E \approx 3.9\% \times E^{-1/2}$, where $E$
is in MeV. The detection efficiency $\varepsilon$ for 5.5-MeV electrons in the BGO crystal was calculated with GEANT4.
The response function consists of Gaussian peak and a flat tail, the number of events in Gaussian peak is $\varepsilon
= 0.67$.


The external $\gamma$ activity was suppressed using a passive shield that consisted of successive layers of lead (90
mm) and bismuth (15 mm ${\rm{Bi}}_2{\rm{O}}_3$). The total thickness of the passive shield was $\approx$ 110 ${\rm{g\;
cm}}^{-2}$.

The setup was located on the Earth's surface. In order to suppress the cosmic-ray background we used an  active veto,
which consisted of five $50\times50\times12$ cm plastic scintillators. The active veto included two energy thresholds:
the first was set at a level $\sim$ 5 MeV, which corresponds to 600 ${\rm{s}}^{-1}$ counts rate and $4\%$ dead time for
70 $\mu\rm{s}$ inhibit pulse. The second stage with $\sim$ 0.1 MeV threshold and $\pm~\rm{6}~\mu\rm{s}$ inhibit pulse
preformed a selection among the BGO impulses for energies exceeding 2.3 MeV. This energy limit was established in order
to make an estimate of the second stage dead time using the 2.614 MeV natural radioactivity peak. Our estimate to the
active shielding dead time is 23\%.

\begin{figure}
\includegraphics[width=9cm,height=10.5cm]{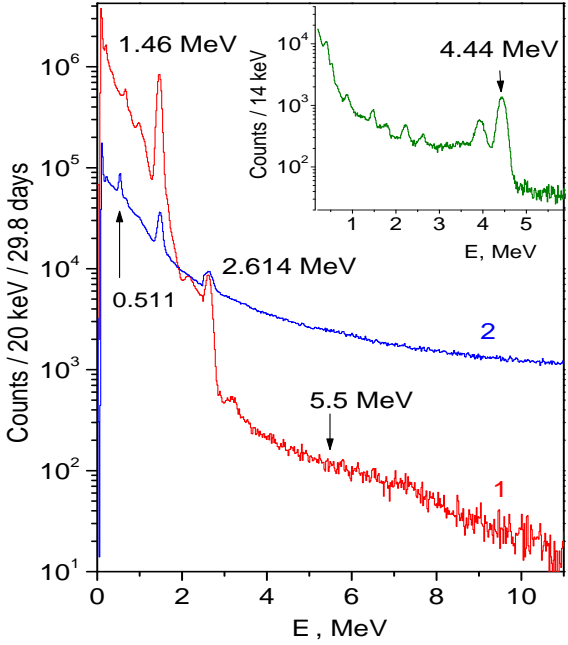}
\caption {The energy spectrum of the BGO detector measured (1) in anticoincidence and (2) in coincidence with  the
active shielding signal. The location of the expected 5.5 MeV axion peak is denoted by an arrow. In inset the spectrum
measured with Pu-Be neutron source is shown.} \label{fig2}
\end{figure}

\section{RESULTS}
The measurements were performed over 29.8 days in live time by 2-hour series. The measurements were divided  into
series in order to monitor the time stability of the BGO detector and the active shielding. The energy spectrum of the
BGO detector in the range of (0--11) MeV is shown in Fig. 2. The spectrum of the BGO signals that were not accompanied
by the active shielding signal is designated as 1.

In the spectrum, one can identify two pronounced peaks at 1.460 MeV and 2.614 MeV; these are due to the natural
radioactivity of the $^{40}{\rm{K}}$ (located in the PMT's glass housing) and of $^{208}{\rm{Tl}}$ from the
$^{232}{\rm{Th}}$ family (Fig. 2).  The positions and intensities of these peaks were used for monitoring of time
stability.

Bismuth has the largest nuclear charge among the stable isotopes (Z = 83), and the cross section of $(e^+e^-)$-pair
production upon the interaction of $\gamma$ quanta is therefore the largest for this element. The annihilation peak at
0.511 MeV is pronounced in the spectrum. The peak at 2.10 MeV is related to the escape of one annihilation $\gamma$
quantum from the detector upon the detection of 2.614 MeV $\gamma$ rays. The visible kink at $\approx$ 7.5 MeV is due
to the $\gamma$ quanta produced as a result of the capture of thermal neutrons by the components of the passive shield.

The positions and dispersion of the 1.46 MeV and 2.614 MeV peaks determined during the measurements were used to  find
the energy scale and resolution of the BGO detector. The energy calibration of the spectrometric channel was found as a
linear function: $E = A\times N + B$, where $E$ is the released energy and $N$ is the channel number. For higher
energies the energy calibration was checked with a $^{239}$Am-$^9$Be neutron source. The reaction
$^9\rm{Be}(\alpha,\rm{n})^{12}\rm{C^*}$ produce $\gamma$-quanta with energy 4.439 MeV corresponding to the energy of
the first exited states $^{12}\rm{C}$ nuclei. The position of 4.439 MeV peak is restored with accuracy better than 5
keV while 1461 keV and 2614 keV calibration peaks are used. (Fig.2, inset).

The dependence of the energy resolution of a scintillation detector vs energy can be written as $\sigma = C\times
\sqrt{E}$.  The parameter C was found to be 0.04 ${\rm{MeV}}^{1/2}$. The values of $\sigma$ determined from the
background spectrum are in good agreement with the measurements performed with $^{60}{\rm{Co}}$, $^{207}{\rm{Bi}}$ and
$^{239}$Pu$^9$Be standard calibration sources. The expected standard deviation of the 5.5 MeV peak due to the axion
absorption is $\sigma =$ 93 keV.

\begin{figure}
\includegraphics[width=9cm,height=10.5cm]{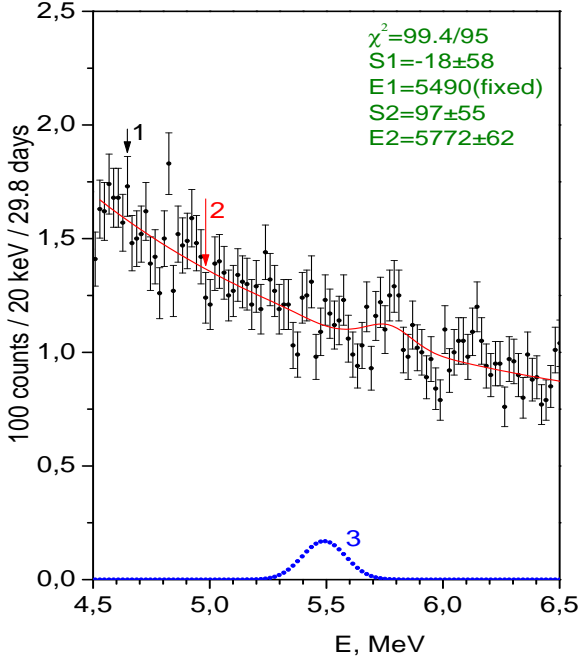}
\caption {The fitted BGO spectrum in the ($4.5-6.5$) MeV range. Curve 3 is the detector response function for  $E_0 =
5.49$ MeV and $\sigma = 0.093$ MeV. } \label{fig3}
\end{figure}

Figure \ref{fig3}  shows the energy range of ($4.5-6.5$) MeV, in which the axion peak was expected. The spectrum
measured in the range of ($4.5-6.5$) MeV was fitted by a sum of exponential and two Gaussian functions:
\begin{equation}
N(E) = a+b\times\exp(cE)+\sum_{i=1}^{2}\frac{S_i}{\sqrt{2\pi\sigma_i}}\exp[-\frac{(E_i-E)^2}{2\sigma_i^2}].
\end{equation}
Here $a$, $b$ and $c$ are parameters of the function describing the smooth background. The position and dispersion of
the first Gaussian peak corresponded to the desired-peak parameters: $E_1 =$ 5.49 MeV is the axion peak position,
$\sigma_1 =$ 0.093 MeV is the Gaussian peak standard deviation.

Because a small unknown peak can be seen at $\approx$ 5.8 MeV, the second Gaussian was added to the fitting function.
The position and area of the second peak were free, while the dispersion $\sigma_2 =$ 0.095 MeV was fixed.

The position of first peak ($E_1$) and dispersion ($\sigma_1$) were fixed and six parameters were varied, three of
which described the continuous background while three others described area of the two peaks ($S_1, S_2$), and the
second peak position ($E_2$). The total number of the degrees of freedom in the range of (4.5--6.5) MeV was 95.

The fit results, corresponding to the minimum $\chi^2$ = 99.4/95, are shown in Fig. \ref{fig3}. The position and  area
of the second peak are: $E_2 = 5.77\pm0.06$ MeV and $S_2 = 97\pm55$ counts. We attribute this peak to the intense 5.824
MeV gamma-ray line resulting in the capture of thermal neutrons  by $^{113}\rm{Cd}$ \cite{NDS05}. The cross section of
thermal neutron capture is $2\times10^4$ barns. The intensity of the 5.49 MeV peak was found to be $S_1 = -18 \pm 58$,
this corresponds to the upper limit on the number of counts in the peak, $S_{lim} = 85$ at a 90\% confidence level
\cite{Fel98}.

 The expected number of axioelectric absorption events are:
\begin{equation}
S_{abs} = \varepsilon N_{Bi}T\Phi_A\sigma_{Ae}
\end{equation}
where $\sigma_{Ae}$ is the axioelectric effect cross section, given by expression (\ref{sigmaAE}); $\Phi_A$ is the
axion flux (\ref{FluxA}); $N_{Bi} = 4.76 \times 10^{24}$ is the number of Bi atoms; $T = 2.57\times10^6$ s is the
measurement time; and $\varepsilon = 0.67$ is the detection efficiency for 5.5 MeV electrons. The axion flux $\Phi_A$
is proportional to the constant $(g^3_{AN})^2$, and the cross section $\sigma_{Ae}$ is proportional to the constant
$g^2_{Ae}$ , according to expressions (\ref{FluxA}) and (\ref{sigmaAE}). As a result, the $S_{abs}$ value depends on
the product of the axion-electron and axion-nucleon coupling constants: $(g_{Ae})^2\times (g^3_{AN})^2$.

\begin{figure}
\includegraphics[width=9cm,height=10.5cm]{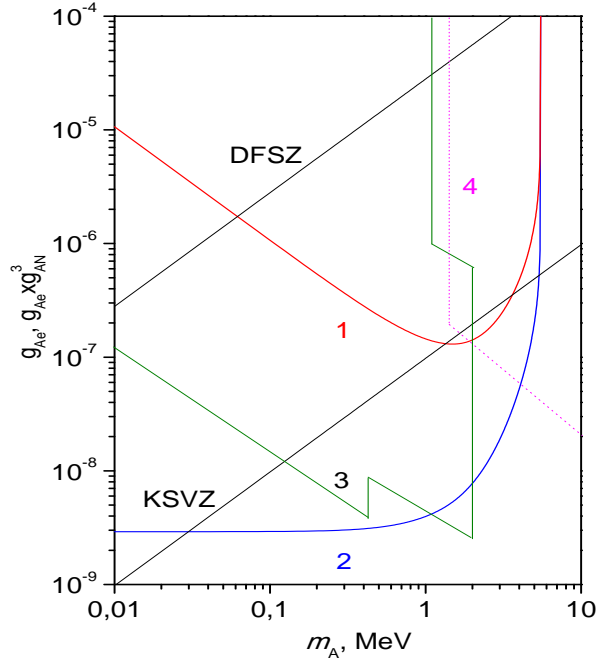}
\caption{The limits on the $g_{Ae}$ coupling constant obtained by 1- present work, 2 - present work for $|g_{Ae}\times
g^3_{AN}|$,  3- solar \cite{Bel08} and reactor experiments \cite{Alt95,Cha07}, 4- beam dump experiments
\cite{Kon86,Bjo88}. The allowed $|g_{Ae}|$ and $|g_{Ae}\times g^3_{AN}|$ values lie below the corresponding curves. The
relations between $g_{Ae}$ and $m_A$ for DFSZ- and KSVZ-models are also shown .} \label{fig4}
\end{figure}

The experimentally found condition $S_{abs} \leq S_{lim}$ imposes some constraints on the range  of possible
$|g_{Ae}\times g^3_{AN}|$  and $m_A$ values. The range of excluded $|g_{Ae}\times g^3_{AN}|$ values is shown in Fig.
\ref{fig4}, at $m_A \rightarrow 0$ the limit is
\begin{equation}
|g_{Ae}\times g^3_{AN}| \leq 2.9\times10^{-9}.
\end{equation}

The dependence of $|g_{Ae}\times g^3_{AN}|$ on $m_A$ is related only to the kinematic factor in formulae (\ref{ratio})
and (\ref{sigmaAE}). These constraints are completely model-independent and valid for any pseudoscalar particle with
coupling $|g_{Ae}|$ less than $10^{-6(4)}$.

Within the hadronic axion model, $g_{AN}^3$  and $m_A$ quantities are related by expression (\ref{gan3}), which can  be
used to obtain a constraint on the $g_{Ae}$ constant, depending on the axion mass (Fig. 4). For $m_A$ = 1 MeV, this
constraint corresponds to $|g_{Ae}|\leq 1.4 \times 10^{-7}$.

Figure 4 also shows the constraints on the constant $|g_{Ae}|$ that were obtained in the Borexino experiment for
478-keV ${^7\rm{Li}}$ solar axions \cite{Bel08} and in the Texono reactor experiment  for 2.2-MeV axions produced in
the $n + p \rightarrow d + A$ reaction \cite{Cha07}. Recently, Borexino coll. reported new more stringent limits on
$g_{Ae}$ coupling for 5.5 MeV solar axions \cite{Bel12}. Unlike our work, these limits on $g_{Ae}$ were obtained in
assumption that the axion interacts with electron through the Compton conversion process.

The sensitivity of the experiment  to the constant $g_{Ae}$ depends on the target mass $M$, specific background level
$B$,  detector resolution $\sigma$, and measurement time $T$: $g_{Ae}^{lim} \sim (\sigma B/MT)^{1/2}$. The level
$g_{Ae}\approx 10^{-10}$ can be achieved with longer measurements with the detector mass enlarged by three orders of
magnitude and the background level reduced by two orders of magnitude, the latter can be done by placing the setup in
an underground laboratory.

\section{CONCLUSIONS}
A search for the axioelectric absorption of 5.5 MeV axions produced in the $p + d \rightarrow {^3{\rm{He}}} + \gamma$
reaction  was performed using a BGO detector with a mass of 2.5 kg, located in a low-background setup equipped with
passive and active shielding. As a result, a model-independent limit on axion-nucleon and axion-electron coupling
constant has  been obtained: $| g_{Ae}\times g_{AN}^3|< 2.9\times 10^{-9}$ (90\% c.l.). Within the hadronic axion model
the constraints on the axion-electron coupling constant $|g_{Ae}|\leq (1.4 - 9.7) \times 10^{-7}$ for axions with
masses $0.1 < m_A < 1$ MeV were obtained for 90\% c.l..

\section{ACKNOWLEDGMENTS}
This work was supported by RFBR grants 13-02-01199-a.

\end{document}